# Pressure induced Superconductivity and location of Fermi energy at Dirac point in $BiSbTe_3$

**Vinod K. Gangwar[a, #], Shiv Kumar[b, #], Mahima Singh[a], Labanya Ghosh[a], Zhang Yufeng[c,d], Prashant Shahi[e], Swapnil Patil[a], Eike F. Schwier[b], Kenya Shimada[b], Yoshiya Uwatoko[c], Sandip Chatterjee[a,*]**

[a]*Department of Physics, Indian Institute of Technology (BHU) Varanasi 221005*
[b]*Hiroshima Synchrotron Radiation Center, Hiroshima University, Higashi-Hiroshima City, 739-0046 Japan*
[c]*Institute for Solid State Physics, University of Tokyo, Kashiwa, Chiba 277-8581, Japan*
[d]*School of Physics and Key Laboratory of MEMS of the Ministry of Education, Southeast University, Nanjing 211189, China*
[e]*Department of Physics, D.D.U. Gorakhpur University, Gorakhpur 273009*

## Abstract

We have grown single-crystal $BiSbTe_3$ 3D TI sample and studied structural, TE as well as pressure dependent magneto-transport properties. Large positive Seebeck coefficient confirmed the *p*-type nature of $BiSbTe_3$, which is consistent with Hall measurement. We have also studied the electronic band structure using Laser-based ARPES, which revealed the existence of a Dirac-cone like metallic surface state in $BiSbTe_3$ with a Dirac Point situated exactly at the Fermi level. Additionally, superconductivity emerges under pressure of 8 GPa with a critical temperature of ~2.5 K. With further increase of pressure, the superconducting transition temperature ($T_c$) increases and at 14 GPa it shows the maximum $T_c$ (~3.3 K).

Topological insulator (TI) is a new quantum state of matter in condensed matter physics which has attracted a huge attention because of their unique properties and potential technological applications such as quantum computation, spintronics and low power-dissipation electronic devices [1-6]. $Bi_2Se_3$, $Bi_2Te_3$ and $Sb_2Te_3$ are most studied 3D TIs, but so far very few reports are available on $BiSbTe_3$TI compounds [7]. Usually, the conductivity of $Bi_2Te_3$ and $Sb_2Te_3$ shows *n*-type behavior and *p*-type behavior respectively due to the formation of Te-vacancies and antisite defects [4, 8-10]. The Fermi level ($E_F$) in *n*-type $Bi_2Te_3$ lies in the bulk conduction band (BCB) and the Dirac point (DP) is hidden in the bulk valence band (BVB). On the other hand, pure $Sb_2Te_3$ has different band structure, in which DP remains within the bulk gap and $E_F$ lies in the BVB (*p*-type). Tuning of the position of DP and $E_F$ can be achieved simultaneously by mixing of these two compounds. This may guide to an ideal TI in which $E_F$ and the DP coincides in insulating bulk, which may lead TIs to the technological applications in a controlled manner. The theoretical and experimental studies on $(Bi_{1-x}Sb_x)_2Te_3$ alloy have shown that the topological properties preserved for the entire range of composition, and with the help of ARPES as well as electrical measurements it was found that a transition from *n*-type to *p*-type was occurred in a composition of 50% (Bi/Sb) [11]. The transition from *n*-type to *p*-type in $(Bi_{1-x}Sb_x)_2Te_3$ films which were grown on sapphire by molecular beam epitaxy (MBE), was about 94% of Sb content [12]. In the other studies, the films of $(Bi_{1-x}Sb_x)_2Te_3$ grown on $SrTiO_3$ substrate show that transition occurs in the range of 35–45% Sb [13]. As a matter of fact, the stoichiometric of Bi/Sb has not yet been reported where $E_F$ is located exactly at the DP. In the present investigation we have shown that in $BiSbTe_3$ 3D TI $E_F$ is located exactly at the DP.

Moreover, $BiSbTe_3$ performs as a good thermoelectric material at room-temperature which attracts a great attention for the application as thermoelectric (TE) devices.

Furthermore, usage of external pressure can be a very useful method to tune the materials properties and design new materials. Because, inter atomic distance inside the material can be reduced by imposing external pressure, which affects the material properties. Recently, pressure-induced superconductivity has been observed in $Bi_2Te_3$, $Bi_2Se_3$ and $Sb_2Te_3$ 3D topological insulators [14-16]. Majorana fermions, potentially fruitful in quantum computing can be expected at the edge states in these 3D topological superconductors [17, 18]. Superconductivity in $Bi_2Te_3$ was observed with the ambient-pressure crystal structure. However, superconductivity in $Sb_2Te_3$ was reported with different crystal structure [14, 16]. The structural transition in $BiSbTe_3$ compound was reported by Jacobsen *et al*. around 8 GPa, which was assigned to the orthorhombic *I222* structure [19]. In their other report, they found a slight increase in Seebeck coefficient and a similar decrease in thermal conductivity at 8 GPa pressure [20].

In this report, we have characterized single crystal of $BiSbTe_3$ sample by ARPES technique to examine the surface states thoroughly and also measured transport properties with the variation of magnetic field and pressure. The single crystal $BiSbTe_3$ 3D TI sample was grown by a modified Bridgman method as has been reported elsewhere [21]. The cleaved shiny silver colored single crystal was characterized by X-Ray diffraction, which indicates that the crystal growth direction is along the *c*-axis as shown in Fig.1 (a). Weyrichet *et al.* estimated Sb content by the intensity ratio of XRD peaks corresponding to (009) and (0015) planes. The Peaks corresponding to (009) and (0012) are absent in $Bi_2Te_3$ but present in $Sb_2Te_3$. Hence, intensity ratio of (009) to (0015) is feasible to determine Sb content. In present sample, intensity ratio is $I_{(009)}/I_{(0015)}$ ~ 0.075, suggested almost 50% of Sb content [22]. We have also performed the powder XRD of $BiSbTe_3$ sample after grinding the single crystal sample. To determine lattice parameters, we have analyzed powder XRD data by Le Bail refinement using the FullProf software (Fig.1. (b)). Determined lattice parameters are $a$=$b$=4.32(4) Å, $c$=30.48(2) Å and cell volume is 493.54(4) $Å^3$ with rhombohedral crystal structure belonging to the space group $D_{3d}^{5}$ (R-3m), whereas the calculated α, β and γ values are 90˚, 90˚ and 120˚. All of the extracted values from Le Bail refinement of XRD spectra are well matched with the reported values [23, 24]. The magneto-transport properties were measured by using physical property measurement system (PPMS) with varying magnetic field up to 7 T. The Seebeck coefficient (S) was measured by homemade setup.

To probe the Dirac point and the surface states of $BiSbTe_3$, the ARPES experiment was performed by Laser-ARPES at the Hiroshima Synchrotron Radiation Center (HiSOR) using s-polarized light of 6.45 eV and measured with a hemispherical analyzer (VG Scienta SES R4000) at 20 K and 60 K. The sample was cleaved in-situ at ~22 K in an ultra high vacuum below $8\times10^{-9}$ Pa.

The cleaved $BiSbTe_3$ single crystal along (*00l*) direction was used for transport properties measurement. The resistivity ($\rho_{xx}$) with respect to temperature (2 K ≤ T ≤ 300 K) is shown in Fig.2(a). From resistivity graph, one can see that the nature of the sample is metallic, as resistivity is decreased with decreasing temperature and become constant at very low temperature. The residual resistivity ratio (RRR) is calculated as $\rho_{xx}$(300 K)/$\rho_{xx}$(2 K) ~9 supporting the good quality of sample.

Moreover, we have performed pressure-dependent measurement on $BiSbTe_3$ single crystal sample and measured electrical resistance (R) as a function of temperature at different pressures. The variation of the resistance (R) at various pressures is shown in Fig. 2(b). From the graph, it can be seen that the value of R decreased when applied pressure changed from 2 GPa to 6 GPa. However, no sharp transition in the resistance was observed up to the lowest measured temperature (~2 K) (inset: Fig. 2 (b)). When pressure was further increased to 8 GPa, a sharp decrease in resistance value is observed around 2.5 K with $T_{onset}$ = 2.47 K. After increasing pressure to 8.25 GPa, a similar transition occurred at a little higher onset temperature ($T_{onset}$ = 2.51 K). This sharp decrease in the resistance at a particular temperature above 8 GPa can be assigned for the superconducting transition. The superconducting transition is getting more pronounced with increasing pressure and $T_{onset}$ is also shifting toward higher value with applied pressure, as shown in the Fig.2 (c). The highest $T_{onset}$ (3.36 K) was observed at 14 GPa, after which it becomes almost constant up to 15 GPa. Fig.2 (c) displayed that above 11 GPa pressure, resistance value dropped to zero before reaching 2 K and simultaneously transition becomes sharper. This superconducting phase of $BiSbTe_3$ can be correlated to the phase transition in $BiSbTe_3$ around 8 GPa as has been reported by Jacobsen *et al.* [19]. Hence, observed superconductivity in $BiSbTe_3$ might be due to the structural phase transition from rhombohedral *R-3m* to orthorhombic *I222*. To confirm the structural transition we measured the resistance as a function of pressure at 300 K as shown in Fig.2 (d, e). The value of resistance decreased with increasing pressure up to 8 GPa. An increase in resistance has been observed above 8 GPa pressure, which decreases slightly in pressure range 8 GPa to 15 GPa. Hence, the increase in resistance is clearly indicating that the structural phase transition occurred around 8 GPa pressure. This is reconfirming that the observed superconductivity phase is due to the structural phase transition in $BiSbTe_3$. Such types of structural and superconducting transitions have also been reported for $Bi_2Te_3$ and $Sb_2Te_3$ around 8 GPa [16, 25]. The variation of temperature at which resistance dropped to zero ($T_C^{zero}$) and $T_{onset}$ with respect to pressure is shown in Fig.2 (d, e).

Magneto-resistance of $BiSbTe_3$ [defined as MR (%) = [{$\rho_{xx}$(B)-$\rho_{xx}$(0)}/$\rho_{xx}$(0)]*100%], as a function of applied magnetic field at different temperatures is shown in Fig.2 (f). The magnetic field was applied in the perpendicular direction to the applied electric current. A non-saturated positive linear MR has been observed at all the temperatures up to 7 T magnetic field; the MR value is ranging nearly from 49% (2 K) to 4% (300 K). Non-saturated MR with high mobility is reported for many topological materials where MR is linearly varied with the carrier mobility [26, 27]. The variation of MR with respect to mobility is shown in Fig.3 (c). From Fig.3 (c) we can see that MR is directly proportional to the mobility in the present case, which is consistent with some other reports [13, 27, 28]. However, the obtained value of mobility and MR from magneto-transport data of present sample is less than those reported by Huang *et al.*[27] but consistent with those reported by other groups [13, 29].

The variation of Hall resistivity ($\rho_{xy}$) with the magnetic field (B) at different temperatures is shown in Fig.3 (a). It can be seen from Fig.3 (a) that Hall resistivity is linear with positive slope in full range of magnetic field and temperature that indicates the transport is *p*-type. The slope of Hall data increases above 100 K; hence the corresponding decrease in the carrier concentration (n) is also observed. The carrier concentration obtained from hall data is ranging from $3.63x10^{19}cm^{-3}$(2 K) to $2.64x10^{19}$ $cm^{-3}$ (300 K) and the obtained mobility from Hall data is ranging from 993 $cm^2$/V.s (2 K) to 164 $cm^2$/V.s (300 K) [13, 24, 29]. It can be observed from the carrier concentration graph that there is a little decrease in carrier concentration below 100 K. The decrease in carrier concentration might be due to the decrease of bulk contribution below 100 K. However, the mobility is decreasing with an increase in temperature; the decrease in mobility with an increase in temperature is mainly because of enhancement of thermal vibration.

The p-type nature of $BiSbTe_3$ TI is also confirmed by thermoelectric measurement with positive seebeck coefficient (*S*), as shown in Fig.3 (d), which is the consequence of contribution of bulk valance band (BVB). The value of '*S*' is increased with temperature and slope is changed above 100 K, which can be correlated with carrier concentration. We obtained the maximum value of '*S*' at room temperature with minimum carrier concentration. The maximum obtained value of Seebeck coefficient is ~240 μV/K at 300 K, which is consistent with other reports with the same carrier density [30-35]. We have calculated the power factor (PF), using the formula PF $=\sigma S^2$, Where σ is the electrical conductivity and *S* is the Seebeck coefficient. The Variation of PF with temperature for $BiSbTe_3$ is shown in inset of Fig.3 (d). It is observed that PF is linearly increased up to 100 K, then saturated at a value ~ $3x10^{-3}W/m.K^2$ [31, 32].

In order to investigate the electronic structure of this compound we have carried out the ARPES measurements as shown in Figs. 4 and 5. These measurements were carried out at 20 K and 60 K respectively. Figs. 4(a) and 4(b) represent the ARPES spectra as a colored and grayscale image respectively. The same holds, respectively, for Figs. 5(a) and 5(b). These images show the BVB of the compound forming a V-shaped valley like bulk band gap region around the Γ-point in the Brilliouin zone and the topological surface state (TSS) residing inside it as observed earlier [11-13]. The sharpness of the linear dispersion of the TSS reveals good quality of the sample as well as the high energy and momentum resolution of the laser ARPES spectra. Quite interestingly we are able to discern the unoccupied part of the Dirac cone (DC) related to the TSS above $E_F$ at 60 K due to the thermal broadening of the Fermi distribution function. The DP is seen to lie near $E_F$ in this compound which happens to be an ideal condition for the technological exploitation of the topological properties of the TSS. This is important because the parent compounds $Bi_2Te_3$ and $Sb_2Te_3$ are mainly of *n*-type and *p*-type [8-13]. Doping Sb for Bi is expected to tune the DP at $E_F$ for some of the intermediate composition. Accordingly, when we doped Sb in place of Bi halfway, we could realize a rare and important feat of tuning the DP to $E_F$. Thus the surface of this particular compound could potentially prove to be useful in realizing the promises of the TSS for practical applications. Surprisingly, the transport properties for this compound show a *p*-type behavior which is difficult to be understood entirely from the states near DP. From our ARPES spectra we observe a finite contribution from the BVB towards $E_F$ which is due to its specific band dispersion enclosing a valley type of bulk band gap containing the DP. Thus when $E_F$ is at the DP the surrounding peaks from the BVB simultaneously also cross $E_F$ giving rise to a net *p*-type electrical transport in this compound since the states of the BVB have a *p*-type dispersion. Nevertheless it is important to investigate further, the methods of tapping the DP towards technological applications.

Figs. 4(c) and 4(d) as well as Figs. 5(c) and 5(d) depict the isoenergy contours and stacked plot of the momentum distribution curves (MDC's) of the ARPES spectra respectively. The stacked plot of MDC's offers clarity in visualizing the DC. Here we are able to see the sharpness of DP at 20 K and the emergence of the unoccupied part of the DC at 60 K clearly. The vertically stacked isoenergy contours taken at different binding energies (BE's) clearly reveal the 3-D view of the DC around the Γ-point. The bright circular central contour surrounding the Γ-point arises from the TSS while the BVB reveals a six-fold petal like structure around the central contour. At higher BE's the central contour touches the petal structure. Upon approaching $E_F$ it shrinks while the petal structure expands which is consistent with the band dispersion of the respective states. Here again we observe a point like Fermi surface (FS) arising due to the DP in concurrence with the occurrence of the DP at $E_F$. At higher BE's the DC is seen to develop slight warping. Such a warping in the FS is known to originate from the effect of spin orbit coupling in the electronic structure of topological insulators [6, 36].

In conclusion, we have investigated the external pressure effect on the resistance and the ARPES on $BiSbTe_3$. It is observed that with increase of pressure resistance decreases and at a value of 8 GPa a sharp drop in resistivity is observed which indicates the occurrence of superconductivity. With further increase of pressure the superconducting transition temperature ($T_c$) increases and at 14 GPa it shows the maximum $T_c$ (~3.3 K). Measured resistivity at 300 K as function of pressure shows an anomaly at 8 GPa indicating the structural phase transition. This may indicate that the superconducting transition might be due to the structural phase transition in $BiSbTe_3$. Furthermore, the ARPES study clearly indicates the location of Fermi energy exactly at Dirac point. This is an ideal condition for the technological exploitation of the topological properties of the TSS. In addition, the large Seebeck coefficient and good TE performance at room-temperature attract great attention for the application in TE devices.

#Both the authors have equal contribution.
*Corresponding author: schatterji.app@itbhu.ac.in

**Figure Captions:**

**Fig.1.** (a) Single crystal x-ray diffraction of $BiSbT_3$ cleaved along (*00l*) direction, Inset: Laue diffraction pattern (b) Le Bail refinement of powder XRD of as grown $BiSbTe_3$ single crystal sample using FullProf software.

**Fig.2.** (a) Resistivity variation with respect to temperature (b) Temperature dependence of resistance at different pressures with a superconducting transition around 8 GPa. (Inset: Variation of resistance with respect to temperature at 2, 4 and 6 GPa pressure) (c) Close view of temperature dependence resistance at different pressures. (d) Pressure dependence of resistance at 300 K with a phase transition around 8 GPa. (e) Variation of superconducting transition temperature with respect to applied pressure of $BiSbTe_3$. (f) Variation of megnetoresistance (MR %) as a function of magnetic field at different temperatures.

**Fig.3.**(a) Displayed magnetic field dependence of the Hall resistivity at different temperatures. (b) Represents the variation of carrier density and mobility as a function of temperature. (c) Shows linear variation of MR with Mobility.

**Fig.4.** (a), (b) ARPES spectra displayed as a colored and a grayscale image at T= 20 K respectively. The stacked plots of (c) the isoenergy contours and (d) the momentum distribution curves (MDC's) of the ARPES spectra at T= 20 K.

**Fig.5.** (a),(b) ARPES spectra displayed as a colored and a grayscale image at T= 60 K respectively. The stacked plots of (c) the isoenergy contours and (d) the momentum distribution curves (MDC's) of the ARPES spectra at T= 60 K.

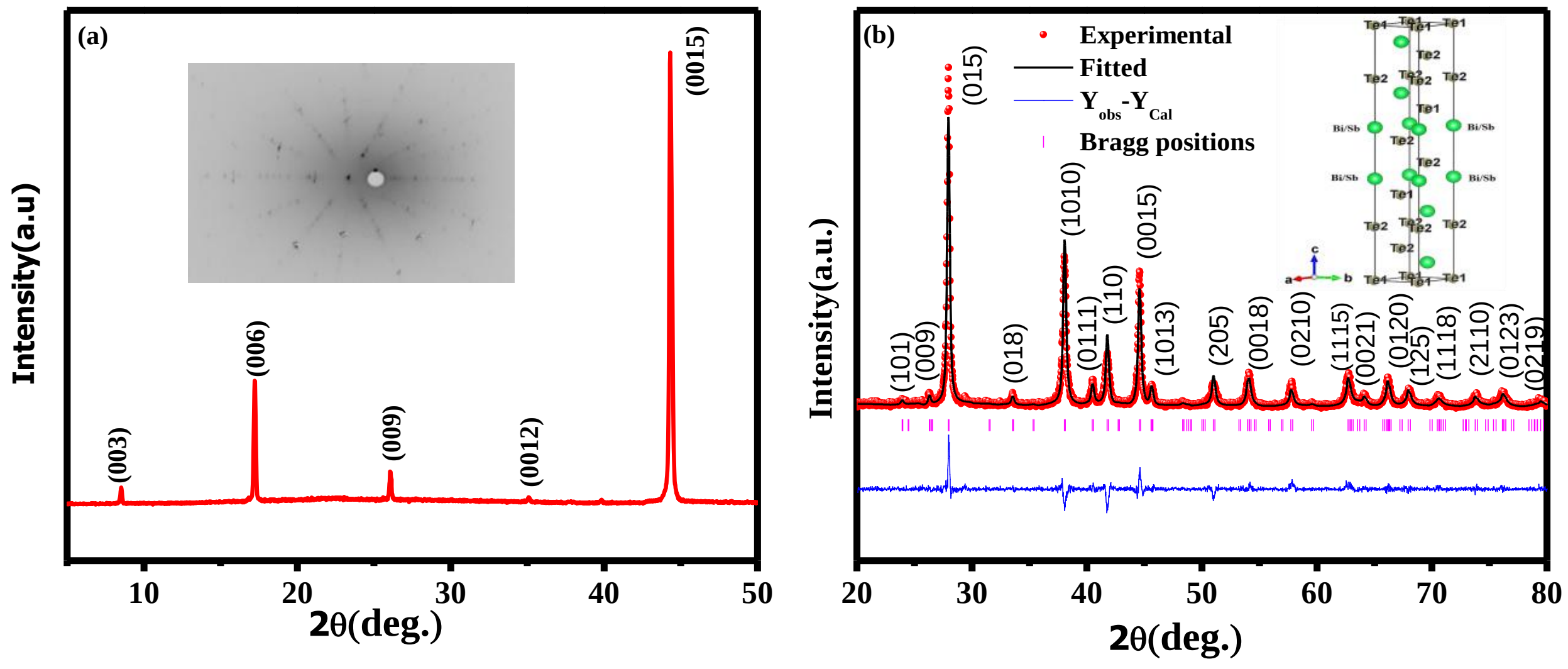

**Fig.1**

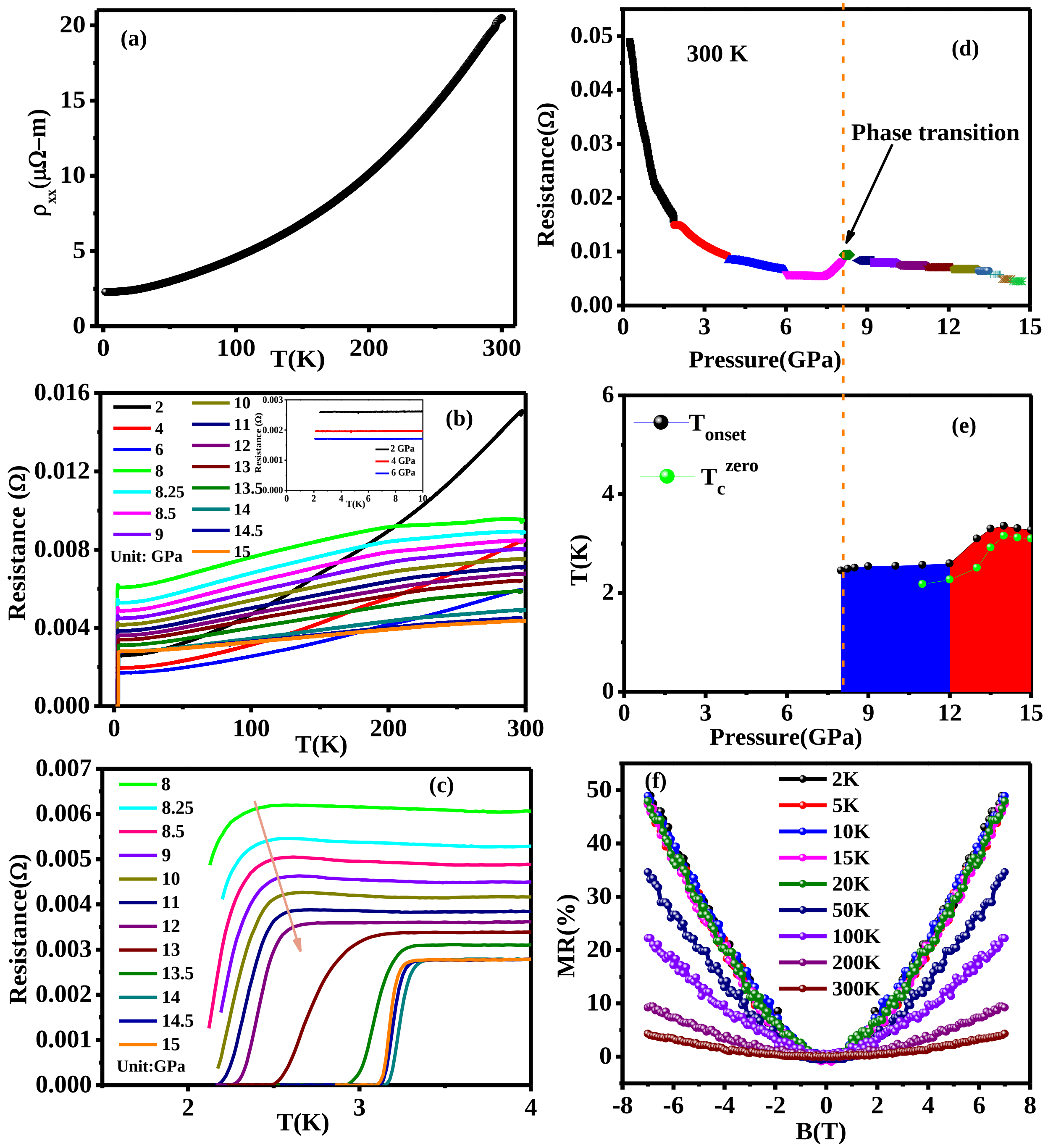

Fig.2

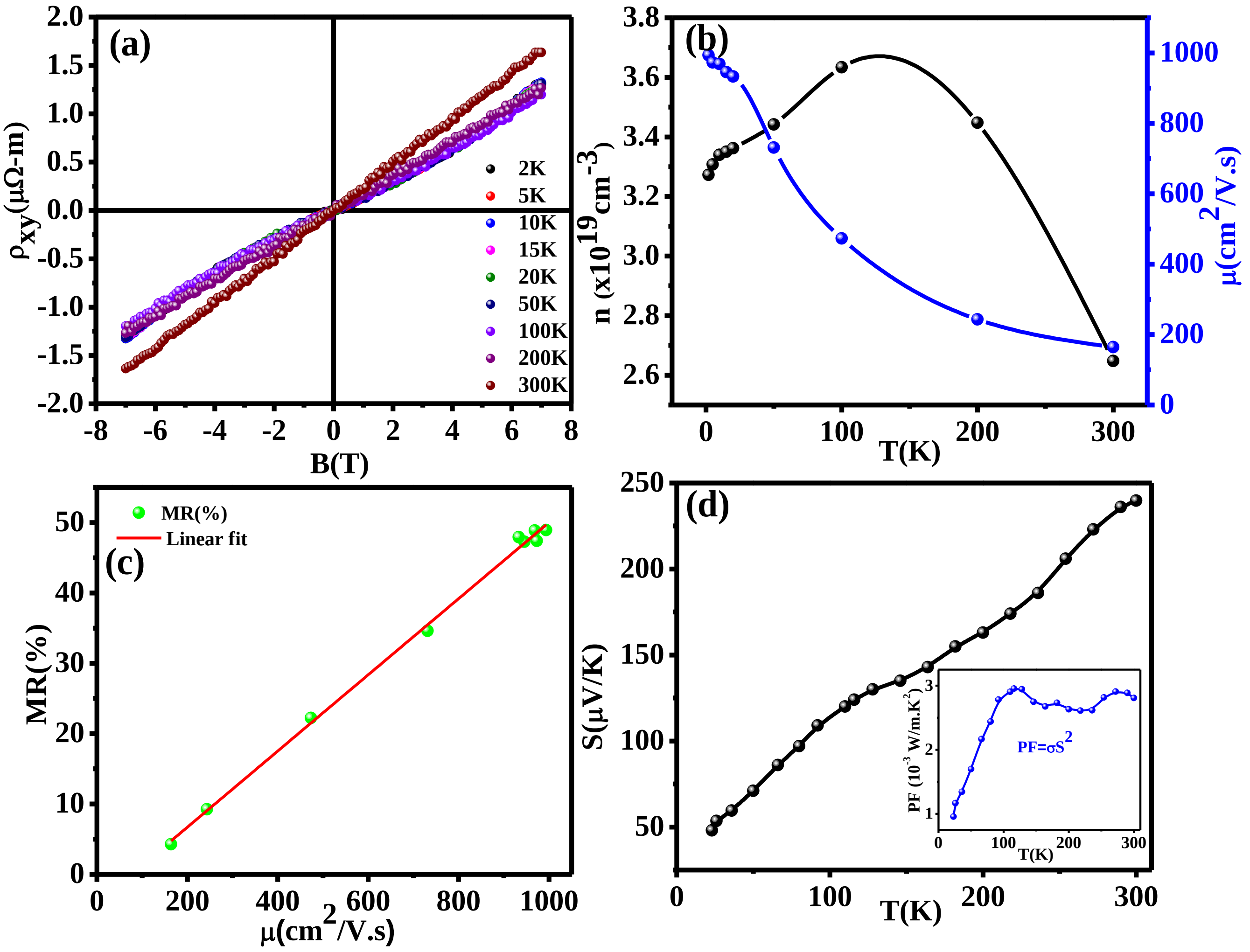

Fig.3

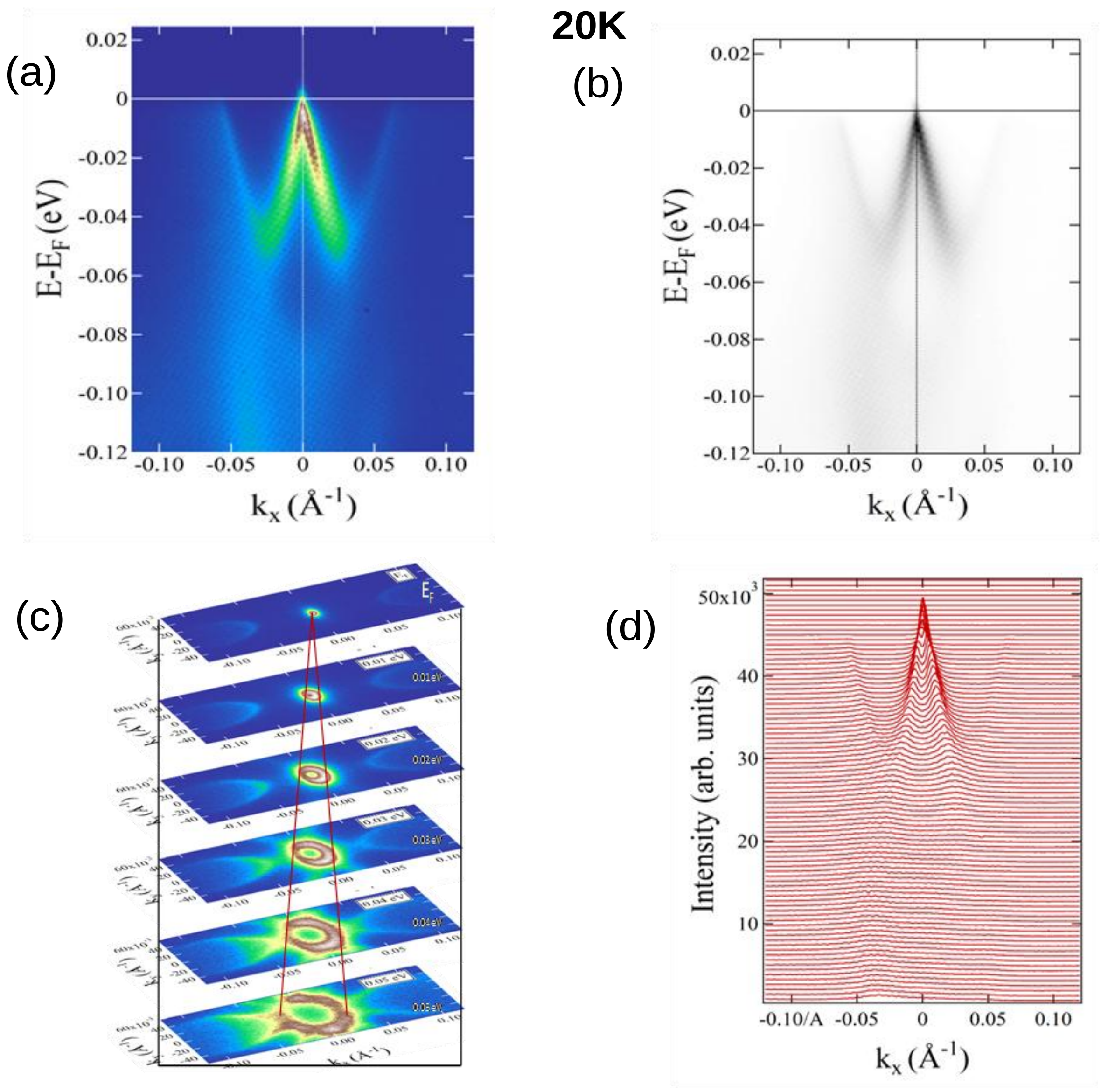

**Fig.4**

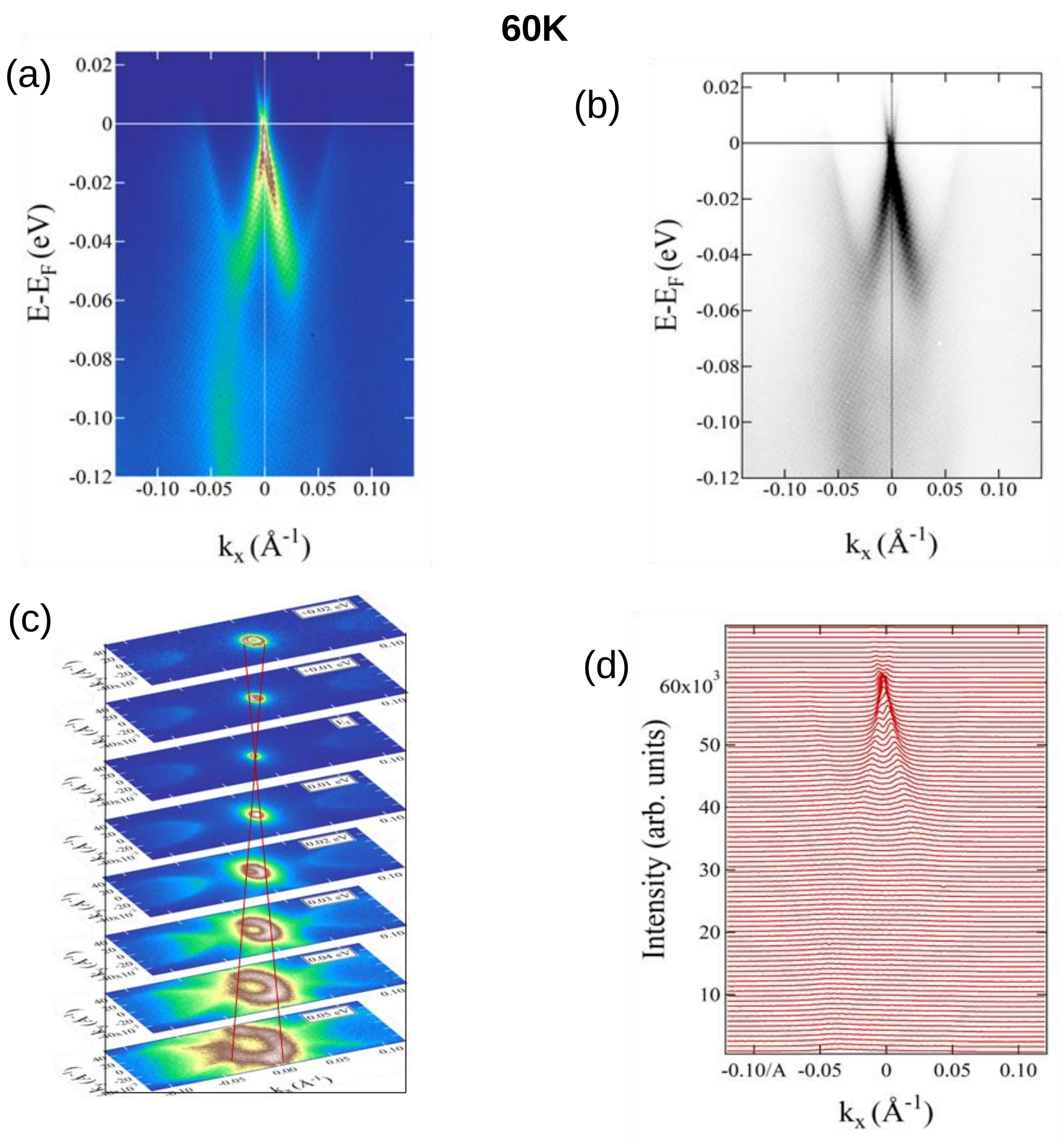

**Fig.5**